\documentclass[ALICE,manyauthors]{cernphprep}
\usepackage{cite,fancyvrb,xspace,graphicx,color}
\usepackage{hyperref}
\usepackage{lineno}

\providecommand*\gevc{GeV/$c$\xspace}

\providecommand{\s}{$\sqrt{s}$\xspace}
\providecommand{\pt}{\ensuremath{p_{\rm T}}\xspace}
\providecommand{\mt}{\ensuremath{m_{\rm T}}\xspace}
\providecommand{\dedx}{d$E$/d$x$\xspace}

\providecommand{\pp}{pp\xspace}

\providecommand{\sigmaTPC}{$\sigma_{\rm{TPC}-\rm{d}\mathit{E}/\rm{d}\mathit{x}}$\xspace}
\providecommand{\sigmaTOF}{$\sigma_{\rm{TOF-PID}}$\xspace}

\begin{document}
\begin{titlepage}
\PHyear{2014}
\PHnumber{097}
\PHdate{15 May}
\title{Measurement of electrons from semileptonic heavy-flavor hadron decays 
       in \pp collisions at $\bf \sqrt{s} = 2.76$~TeV}
\ShortTitle{Electrons from heavy-flavor decays in \pp collisions at 
            \mbox{\s$ = 2.76$~TeV}}
\Collaboration{ALICE Collaboration%
  \thanks{See Appendix~\ref{app:collab} for the list of collaboration
    members}}
\ShortAuthor{ALICE Collaboration}
\begin{abstract}
The \pt-differential production cross section of electrons from semileptonic 
decays of heavy-flavor hadrons has been measured at mid-rapidity 
in proton-proton collisions at \mbox{\s$ = 2.76$~TeV} 
in the transverse momentum range \mbox{0.5~$<$ \pt $<$~12~GeV/$c$} with
the ALICE detector at the LHC. The analysis was performed using minimum bias 
events and events triggered by the electromagnetic calorimeter. Predictions 
from perturbative QCD calculations agree with the data within the theoretical 
and experimental uncertainties.
\end{abstract}
\end{titlepage}
\setcounter{page}{2}


\section{Introduction}
\label{sec:intro}

The measurement of the production of heavy-flavor hadrons, i.e. hadrons 
carrying charm or beauty quarks, in proton-proton (pp) collisions provides 
a test of quantum chromodynamics (QCD), the theory of the strong interaction. 
In hadronic collisions, heavy quarks are almost exclusively produced through 
initial hard partonic scattering processes because of their large 
masses~\cite{Lourenco:2006vw}. 
Consequently, the heavy-flavor hadron production cross sections are calculable 
in the framework of perturbative QCD (pQCD) down to very low transverse momenta 
(\pt).

Furthermore, heavy-flavor production cross sections measured in pp collisions 
provide a reference for corresponding measurements in high-energy 
nucleus-nucleus collisions, in which the formation of a strongly interacting 
partonic medium has been observed~\cite{Averbeck:2013oga,Arsene:2004fa,Adcox:2004mh,Back:2004je,Adams:2005dq,Abelev:2012eq,CMS:2012aa,Aad:2010bu}. 
Heavy quarks are produced on short timescales, presumably before this medium 
is formed. Consequently, they probe the medium properties while they propagate 
through it~\cite{Adare:2006nq,Abelev:2006db,ALICE:2012ab,Abelev:2012qh}. In 
particular, the color charge and mass dependence of the partonic energy loss 
can be studied by comparing the suppression of heavy-flavor hadrons and hadrons
carrying light quarks only~\cite{Dokshitzer:2001zm,Wicks:2007am}.

One available method to investigate heavy-flavor production is the measurement 
of the contribution of semileptonic decays of heavy-flavor hadrons to the 
inclusive electron spectra. This contribution is substantial because of 
branching ratios of the order of 10\% into the semielectronic decay 
channel~\cite{Beringer:1900zz} and the large heavy-quark production 
cross sections at LHC energies \cite{ALICE:2011aa,Abelev:2012sca}. 
In \pp collisions at \mbox{\s = 7 TeV}, the signal of electrons from 
heavy-flavor hadron decays is of similar magnitude as the 
background~\cite{Abelev:2012xe} at an electron transverse momentum of 
$\approx 2$~\gevc, and the ratio of signal to background increases with \pt.

The production of heavy-flavor hadrons was studied at the LHC in \pp collisions
at \mbox{\s = 7 TeV} in various channels by 
ALICE~\cite{Abelev:2012xe,Abelev:2012sca,Abelev:2012pi,ALICE:2011aa,Abelev:2012gx}, 
ATLAS~\cite{Aad:2011rr,Aad:2012fc,Aad:2011sp}, 
CMS~\cite{Khachatryan:2010yr,Chatrchyan:2012xg,Chatrchyan:2012hw,Chatrchyan:2011vh,Chatrchyan:2011pw,Khachatryan:2011hf,Khachatryan:2011mk}, 
and LHCb~\cite{Aaij:2012jd,Aaij:2010gn,Aaij:2011jh,Aaij:2013mga}. 
Perturbative QCD calculations~\cite{Cacciari:2012ny,Kniehl:2012ti,Bolzoni2013253,Bolzoni2013334,Maciula:2013wg} 
describe the measurements within the uncertainties.

For a center-of-mass energy of 2.76 TeV, which is the reference energy for 
Pb--Pb collisions in 2010 and 2011 at the LHC, ALICE already reported on the 
production of muons from heavy-flavor hadron decays in \pp collisions at 
forward rapidity~\cite{Abelev:2012qh}, and reconstructed open charm mesons at 
mid-rapidity~\cite{Abelev:2012sx}. Again, pQCD calculations describe the 
experimental data reasonably well. This paper presents a measurement of 
electrons, \mbox{(e$^+$+e$^-$)/2}, from semileptonic decays of charm and 
beauty hadrons in the transverse momentum range 
\mbox{0.5~$<$ \pt $<$~12~GeV/$c$} at mid-rapidity in \pp collisions at 
\mbox{\s$ = 2.76$~TeV} using the ALICE detector. The analysis technique 
employed here is similar to the one described in detail in 
\cite{Abelev:2012xe}, where the measurement in \pp collisions at 
\mbox{\s = 7 TeV} is presented, and it consists of the following steps: 
selection of electron candidates, subtraction of the remaining hadron 
contamination, correction for efficiency and normalisation, and subtraction 
of the electron background originating from non-heavy-flavor sources.

\section{Experimental setup and dataset}
\label{sec::Dataset}
The ALICE experiment at the LHC is described in detail in~\cite{ALICE}, thus 
we only briefly introduce the detectors relevant for this analysis.

The detector closest to the interaction point is the Inner Tracking System 
(ITS). It consists of six cylindrical layers, grouped into three subsystems. 
The Silicon Pixel Detector (SPD) equips the two innermost layers, placed at 
radii of 3.9~cm and 7.6~cm from the beam axis. The spatial resolution of the 
detector is 12~$\mu$m in the transverse plane ($r\varphi$) and 100~$\mu$m 
along the beam direction. The SPD is followed by two layers of the Silicon 
Drift Detector (SDD) and two layers of the Silicon Strip Detector at radii 
between 15~cm and 43~cm.

A large cylindrical Time Projection Chamber (TPC), which is the main tracking 
detector, surrounds the ITS at a radial distance between 85~cm and 247~cm. 
The chamber's volume is filled with a mixture of  Ne (85.7\%), CO$_2$ (9.5\%), 
and N$_2$ (4.8\%) as drift gas. In the radial direction, the readout is divided
into 159 pad rows. The TPC covers a pseudorapidity range of $|\eta| < 0.9$ for 
tracks having space points in the outermost pad rows. The specific energy 
deposit \dedx is used to identify particles. The \dedx resolution of the TPC 
(\sigmaTPC) is approximately 5.5\% for minimum ionizing particles passing 
through the full detector~\cite{thesis_kalweit}. 

The tracking detectors are housed inside a solenoidal magnet providing a 
homogeneous magnetic field of 0.5~T. The ITS and the TPC provide a transverse 
momentum measurement for charged particles with a resolution of $\approx$1\% 
at 1~\gevc and $\approx$3\% at 10~\gevc \cite{Aamodt:2010my}.

The Time-of-Flight detector (TOF) is located at a distance of 3.7~m from the 
beam axis covering the full azimuth and $|\eta| < 0.9$. The resolution of the 
particle arrival time is better than 100~ps. The collision time $(t_{0}$) is 
measured with the T0 detector, an array of Cherenkov counters positioned at 
$+370$~cm and $-70$~cm, respectively, along the beam axis. In case no 
information from the T0 detector is available, the collision time is estimated 
using the arrival time of the particles in the TOF detector. If also this 
second method does not provide a $t_{0}$ measurement, the bunch crossing time 
from the LHC is used~\cite{Abelev:2012sx}. Particles are identified using 
the difference between the measured time-of-flight and the expected 
time-of-flight for a given particle species, normalized to the overall 
time-of-flight resolution \sigmaTOF $\approx 150$~ps~\cite{Abelev:2012sx}, 
including both the resolution of the particle arrival time measurement and 
of the $t_{0}$.

The Electromagnetic Calorimeter (EMCal) is a sampling calorimeter based on 
Shashlik technology spanning the pseudorapidity range \mbox{$|\eta|<0.7$} and 
covering $107^{\circ}$ in azimuth~\cite{Abelev:2013fn}. The azimuthal coverage 
was limited to $100^{\circ}$ for the data presented here. The EMCal supermodules
comprise individual towers each spanning 
\mbox{$\Delta\varphi\times\Delta\eta = 0.0143\times0.0143$} (6 $\times$ 6~cm). 
Each $2 \times 2$ group of neighboring EMCal towers forms a trigger elementary 
patch. The energy resolution was measured to be 
1.7 $\bigoplus$ 11.1/$\sqrt{E({\rm GeV})}$ $\bigoplus$ 5.1/$E({\rm GeV})\%$~\cite{Allen:2009aa}, 
where $\bigoplus$ indicates a sum in quadrature.

The V0 detector, used for online triggering and offline event selection, 
consists of two arrays of 32 scintillator tiles on each side of the interaction
point. The detectors cover \mbox{$2.8 < \eta < 5.1$} and 
\mbox{$-3.7 < \eta < -1.7$}, respectively. 

The data used in this analysis were recorded in spring 2011. Two different 
data samples are available: a minimum bias sample and a sample triggered by 
the EMCal. In both samples, the SDD information was read out only for a 
fraction of the recorded events. The minimum bias trigger required at least one
hit in either of the V0 detectors or the SPD. Background from beam-gas 
interactions was eliminated using the timing information from the V0 
detector and the correlation between the number of hits and the reconstructed 
track segments in the SPD~\cite{Aamodt:2009aa}. Events were required to have a 
reconstructed primary vertex~\cite{Aamodt:2010my} within \mbox{$\pm$10 cm} 
from the center of the detector along the beam direction. This covers 86\% of 
all interactions. Pile-up events were identified as events having multiple 
vertices reconstructed in the SPD and they were rejected in this analysis. 
The probability of pile-up events was less than 2.5\% in this data sample. 
The amount of remaining pile-up events after rejection was negligible in this 
analysis~\cite{Abelev:2012xe}. Before further event selection the minimum bias 
sample consisted of 65.8 M events, corresponding to an integrated luminosity 
$L_{\rm int} = 1.1~\rm{nb}^{-1}$. The use of the TOF information for particle 
identification required a stricter run selection which limited the integrated 
luminosity to 0.8~$\rm{nb}^{-1}$ (43.8 M events). In addition to the minimum 
bias sample, events selected by the EMCal trigger were analyzed. It required 
the coincidence of the minimum bias trigger condition described above and an 
energy sum in $2 \times 2$ EMCal trigger patches ($4 \times 4$ towers) 
exceeding nominally 3~GeV. After event selection, the data sample recorded 
with the EMCal trigger corresponded to an integrated luminosity of 
$L_{\rm int} = 12.9$~nb$^{-1}$.

\section{Analysis}
\label{sec::Analysis}

The minimum bias data sample was analyzed employing electron identification
based on the information from the TPC~\cite{Alme:2010ke}. At low transverse 
momentum ($\pt < 2$~\gevc) additional information from the TOF detector was 
required to improve the rejection of hadronic background. Electron 
identification in the analysis of the EMCal triggered data sample was based on 
the combined information from the TPC and the EMCal. The three analyses 
employing TPC, TPC-TOF, and TPC-EMCal electron identification, were conducted 
in different kinematical regions. In transverse momentum, the TPC analysis was 
restricted to the range $2 < \pt < 7$~\gevc, the TPC-TOF analysis was performed
in the range $0.5 < \pt < 5$~\gevc, and the TPC-EMCal analysis was done in the 
range $2 < \pt < 12$~\gevc. In the latter case, the analysis used the minimum 
bias data sample for electron transverse momenta below 5~\gevc and an EMCal 
triggered data sample for electron \pt above 4~\gevc. 
In \pt regions where the cross sections have been determined from more than 
one analysis the results were found to be consistent within uncertainties. 
Results from individual analyses were adopted for three different \pt ranges.
At low \pt (up to 2~\gevc), the TPC-TOF analysis provides the purest electron
candidate sample. In the range \mbox{2~$<$ \pt $<$~4.5~GeV/$c$}, the result 
from the TPC analysis has smaller systematic uncertainties than both results
from the TPC-TOF and the TPC-EMCal analyses. At high \pt (above 4.5~\gevc), 
the TPC and TPC-TOF analyses are statistics limited and the TPC-EMCal analysis 
of the EMCal triggered data sample provides the smallest uncertainty.

\begin{table}[ht]
\label{tab::trackcuts}
\begin{center}
\begin{tabular}{l|c|c}
\hline\hline
Analysis & TPC-TOF/TPC & TPC-EMCal \\
\pt range (\gevc)& 0.5 -- 4.5 & 4.5 -- 12 \\
\hline
Number of ITS clusters &  $\geq$ 3 & $\geq$ 3 \\
SPD layer in which a hit is requested & both & any \\
Number of TPC clusters &  $\geq$ 120 & $\geq$ 120 \\
Number of TPC clusters in \dedx calculation & $\geq$ 80 & - \\
Distance of closest approach to the prim. vertex in $xy$ & $<$ 1 cm & $<$ 1 cm\\
Distance of closest approach to the prim. vertex in $z$ & $<$ 2 cm & $<$ 2 cm \\
$\chi^{2}$/ndf of the momentum fit in the TPC & $\leq$ 4 & $\leq$ 4\\
Ratio of found/findable TPC clusters~\cite{thesis_kalweit} & $\geq$ 0.6 & $\geq$ 0.6 \\
\hline\hline
\end{tabular}
\end{center}
\caption{Summary of the track selection cuts utilized in the different 
         analyses. The same track selection cuts are applied in the TPC-TOF 
         and the TPC analyses.}
\end{table}

Reconstructed tracks were selected for the analysis using the criteria 
listed in Table~\ref{tab::trackcuts}, which are similar to those used in the 
analysis described in ~\cite{Abelev:2012xe}. In particular, the cut on the 
minimum number of ITS clusters was reduced to three (instead of the value of 
four used in~\cite{Abelev:2012xe}) because the SDD points, which were not 
available for a sizeable fraction of the events, were excluded from the track 
reconstruction used for this analysis, thus limiting the maximum number of 
hits in the ITS to four. In order to reduce wrong associations between 
candidate tracks and hits in the first layer of the SPD, hits in both layers 
of the SPD were required in the TPC-TOF analysis. In the TPC-EMCal analysis, 
this requirement has been relaxed to at least one hit in any of the two SPD 
layers in order to increase the statistics, thus resulting in a larger 
background. A cut on the minimum distance to the primary vertex was not imposed
because electrons from charm hadron decays are indistinguishable from 
electrons originating from the primary vertex. 

Three methods were used to identify electrons: in both the TPC and the TPC-TOF 
analyses, electrons were identified via their specific energy deposition 
(\dedx) in the TPC. Tracks were required to have a \dedx between one standard 
deviation below and three standard deviations above the expected \dedx of 
electrons, consistent with an electron identification efficiency of 
$\approx 85$\%. In the TPC analysis for \pt $\geq$ 2 GeV/$c$, a more stringent 
cut was applied in order to cope with the increasing hadron contamination 
towards higher momenta. Therefore, electron candidate tracks were required to 
have a \dedx between 0.5 standard deviations below and three standard 
deviations above the mean \dedx for electrons, corresponding to a selection 
efficiency of $\approx 70$\%. For \pt $<$ 2 GeV/$c$, in the TPC-TOF analysis, 
the TOF detector was used in addition to the TPC. Here tracks were required to 
have a time-of-flight consistent with the expected time-of-flight for electrons
within 3 standard deviations \sigmaTOF, thus rejecting protons and kaons at 
momenta where they cannot be distinguished from electrons via \dedx alone. 

\begin{table}
\label{tab::sysuncertainty}
\begin{center}
\begin{tabular}{l|c|c}
\hline\hline
Analysis & TPC-TOF/TPC & TPC-EMCal \\
\pt range & 0.5 -- 4.5~\gevc & 4.5 -- 12~\gevc \\
\hline
ITS-TPC matching & 2\% & 2\% \\
ITS clusters & 3\% & 3\% \\
TPC clusters & 2\% & 3\% \\
TPC clusters for PID & 2\% & 2\% \\
DCA & negligible & negligible \\
Unfolding & 1\% & 2\% \\
TOF PID & \pt $<$ 2 GeV/$c$: 2\% & -- \\
TPC PID & \pt $<$ 4.5 GeV/$c$: 2\% & -- \\
TPC-EMCal PID & -- & \pt = 4.5 GeV/$c$: 10\%\\
              &    & \pt = 12 GeV/$c$: 20\% \\
Trigger rejection factor & -- & 3\% \\
Rapidity and charge & 2\% & 2\% \\
\hline\hline
\end{tabular}
\end{center}
\caption{Contributions to the systematic uncertainties on the inclusive 
         electron spectrum for the different analyses.}
\end{table}

For $\pt \geq 4.5$~\gevc, the TPC-EMCal analysis was employed. In the TPC, a 
\dedx between 1.4 standard deviations below and three standard deviations 
above the mean \dedx for electrons was required, corresponding to an electron 
identification efficiency of $\approx 90\%$. Tracks were extrapolated from the 
TPC to the EMCal surface and geometrically associated with EMCal clusters 
within 0.02 both in $\eta$ and in $\varphi$. The ratio of the energy of the 
matched cluster in the EMCal to the momentum measured with the TPC and ITS 
($E/p$) was required to be within 0.8 and 1.4 for electron candidates, 
corresponding to an identification efficiency of $\approx 60\%$ averaged 
over \pt. 

The hadronic background was estimated using a parameterization of the TPC 
\dedx in various momentum slices~\cite{Abelev:2012xe} or, alternatively, 
the $E/p$ distribution of identified hadrons, and it was subtracted from 
the electron candidate sample. For the TPC-TOF/TPC analysis the hadron 
contamination was negligible for $\pt \leq 2$~\gevc and less than 1.5\% for 
$\pt \leq 4.5$~\gevc. In the TPC-EMCal analysis, the hadron contamination 
was negligible for $\pt \leq 6$~\gevc, remained below 10\% for 
$\pt \leq 8$~\gevc, and it increased to $\approx 40$\% at $\pt = 12$~\gevc.

The \pt-differential invariant yield of inclusive electrons per minimum bias 
event has been obtained by dividing the raw yield of electrons, 
\mbox{(e$^+$+e$^-$)/2}, measured in \pt bins of widths $\Delta\pt,$ by the 
number of minimum bias events, by $2\pi\pt^{\rm{center}}$ where $\pt^{\rm{center}}$
is the value of \pt at the center of each bin, by $\Delta\pt$, by the width 
$\Delta y$ of the covered rapidity interval, and by the product of the 
geometric acceptance $\epsilon^{\rm geo}$, the reconstruction efficiency 
$\epsilon^{\rm reco}$, and the electron identification efficiency 
$\epsilon^{\rm eID}$. In the TPC-TOF/TPC analyses, $\epsilon^{\rm geo}$, 
$\epsilon^{\rm reco}$, and $\epsilon^{\rm eID}$ in TOF were obtained using a 
Monte Carlo simulation. Proton-proton events at \s = 2.76 TeV were 
generated with the PYTHIA 6.4.21 event generator~\cite{Sjostrand:2006za}. 
Two samples were used for the efficiency calculation: a minimum bias sample 
based on the Perugia-0 tune~\cite{Skands:2009zm} and a heavy-flavor enhanced 
sample containing only events with at least one c$\bar{\rm c}$ or 
b$\bar{\rm b}$ pair. 
The enhanced sample provided a sufficient number of tracks for efficiency 
determination in the \pt region above 4~\gevc. Tracks were propagated 
through the detector using GEANT3~\cite{Geant3}. The electron selection 
efficiency in the TPC ($\epsilon_{ID}$) was extracted from data using the 
measured mean \dedx and the width of the \dedx distribution for electrons. 
The product of acceptance and efficiency was $\approx 0.3$, with a mild 
dependence on \pt. In the TPC-EMCal analysis, the reconstruction efficiency 
was obtained in a similar way to the TPC-TOF/TPC analyses, and the electron 
selection efficiency was determined again from data utilizing the measured 
mean \dedx.

\begin{figure}[b]
  \centering
  \includegraphics[width=0.5\textwidth]{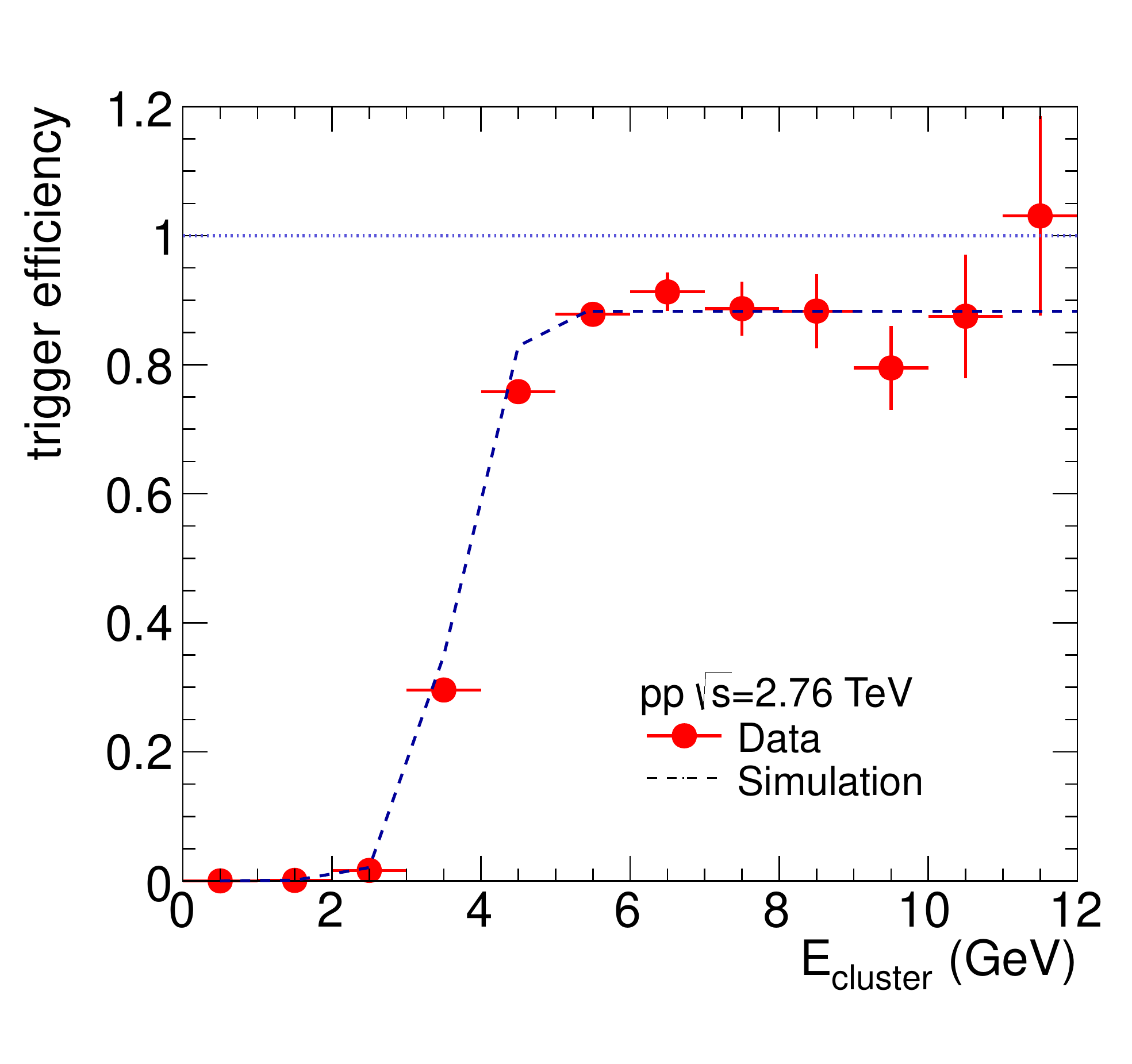}
  \caption{Efficiency of the EMCal trigger as a function of the cluster
           energy measured in the calorimeter~\cite{Abelev:jetpp}.}
  \label{Fig::triggercurve}
\end{figure}

In addition, a correction for the trigger bias was applied in the EMCal 
triggered data sample. This correction was determined from the ratio of the 
EMCal cluster energy distribution in triggered data compared to those in 
minimum bias data. The resulting rejection factor at high energy  
(above the nominal trigger threshold of 3~GeV) was determined to be 
1180$\pm$10. The trigger efficiency is shown in Fig.~\ref{Fig::triggercurve} 
as a function of the cluster energy~\cite{Abelev:jetpp}. The trigger efficiency
obtained from data is well-reproduced by a simulation which incorporated the 
supermodule-by-supermodule variation in the trigger turn-on curves and took 
into account the trigger mask employed in data. The statistics of the minimum 
bias data sample were such that a precise measurement of the trigger efficiency
for electrons as a function of track $\pt$ was not possible. Thus, the 
trigger simulation was used to generate a trigger efficiency for 
electrons as a function of track $\pt$. Above 5~GeV the trigger efficiency is 
$\approx$~85\%, limited by the trigger mask. 

The precision of the transverse momentum measurement is limited by the 
momentum resolution and it is affected by the energy loss of electrons via 
Bremsstrahlung in material. To correct for the resulting distortion of the 
shape of the inclusive electron \pt distributions, an unfolding procedure 
based on Bayes' theorem~\cite{DAgostini:1994zf} was used. 

In order to evaluate the systematic uncertainty, the analysis was repeated 
with modified track selection and particle identification criteria. 
Table~\ref{tab::sysuncertainty} gives an overview of the systematic uncertainty 
assigned to various contributions. The total systematic uncertainty of the 
TPC-TOF/TPC analysis is less than 6\% for \pt~$< 4.5$~\gevc. The systematic 
uncertainty of the TPC-EMCal analysis grows from 10\% at \mbox{4.5 \gevc} to 
20\% at \mbox{12 \gevc}.

\begin{figure}
\includegraphics[width=0.48\textwidth]{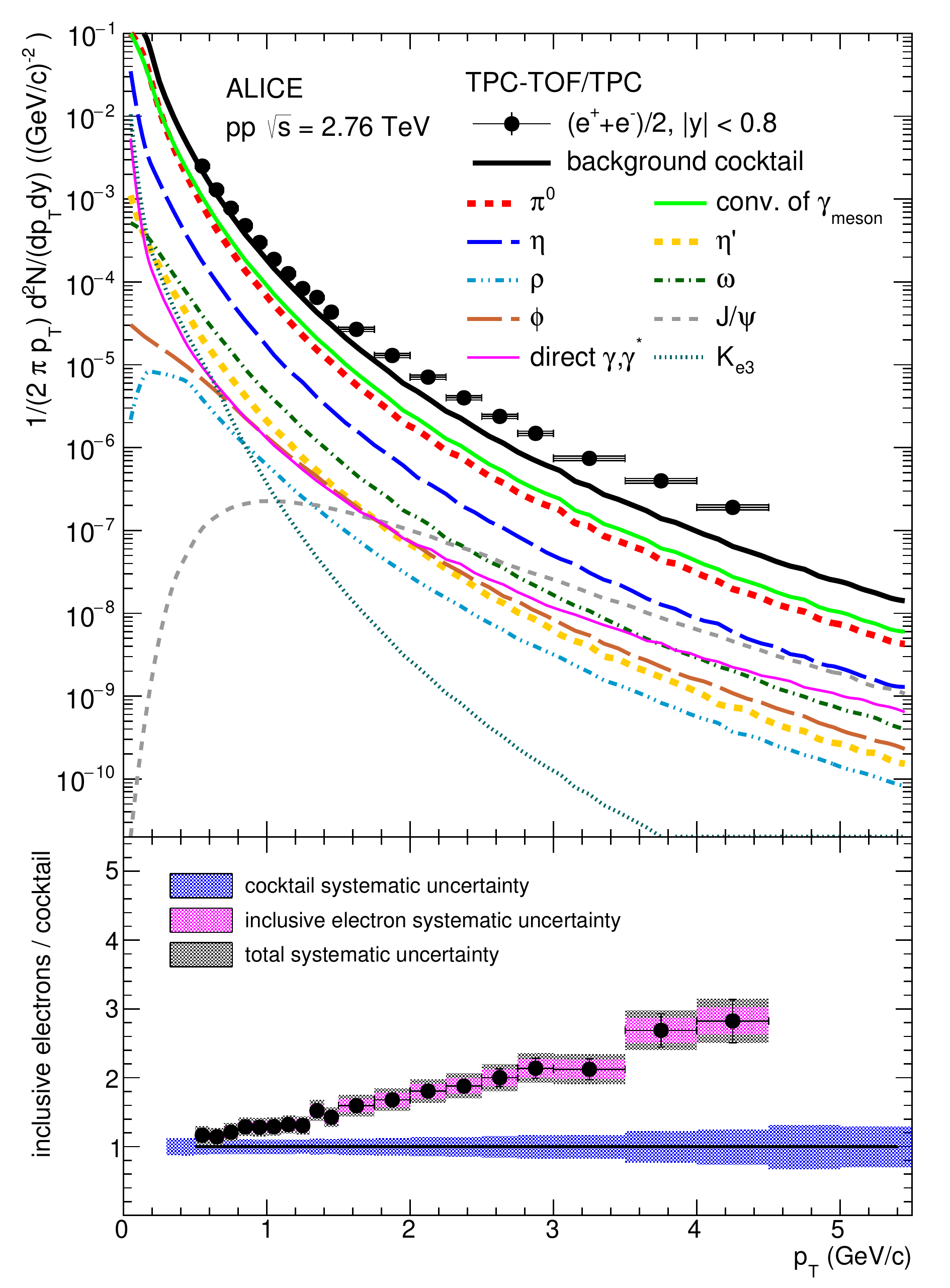}
\includegraphics[width=0.48\textwidth]{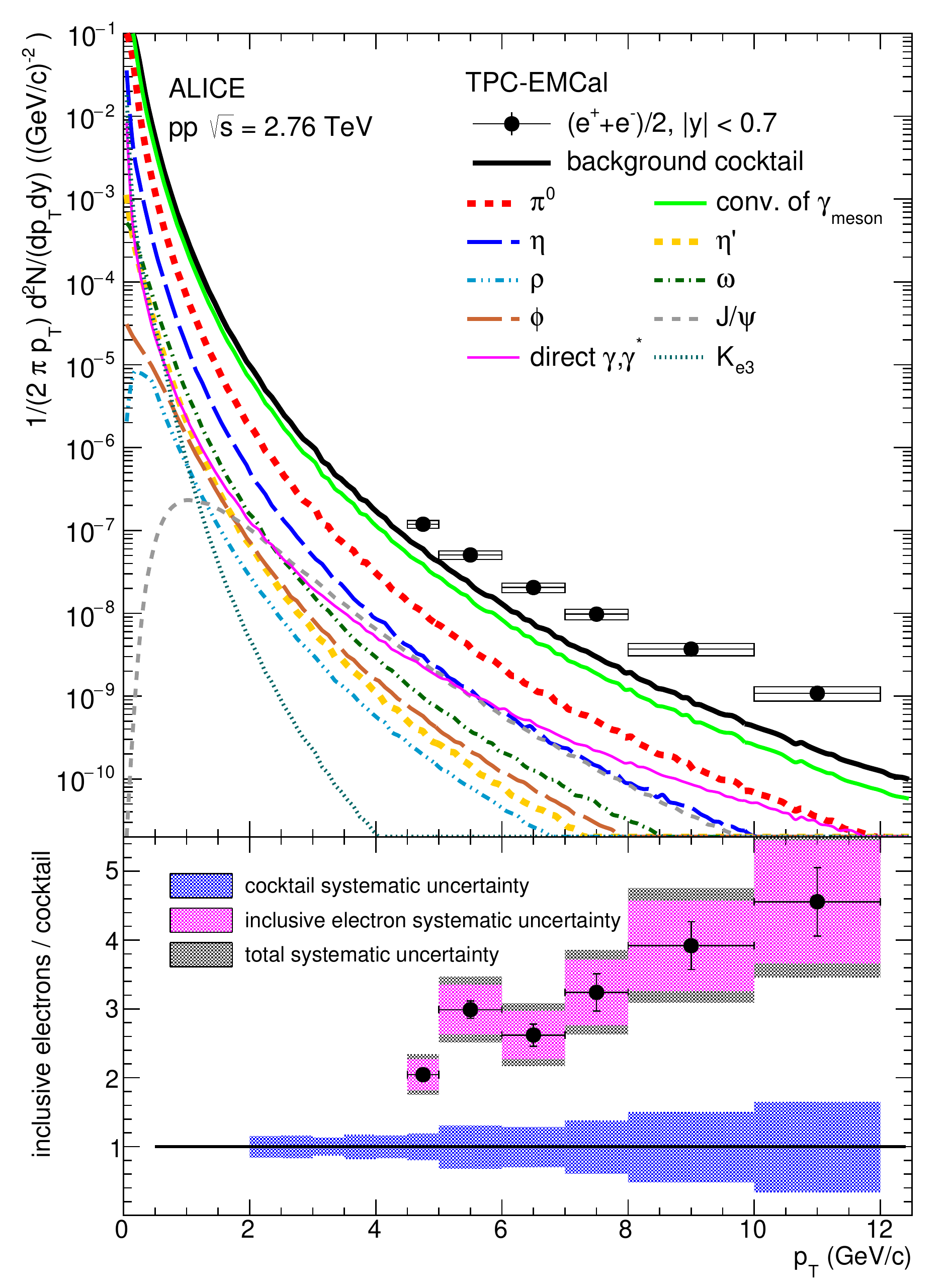}
\caption{\pt-differential invariant yield of inclusive electrons compared to
the electron background cocktail for the TPC-TOF/TPC analysis (left) and the 
TPC-EMCal analysis (right). Ratios of the inclusive electron yields to the
respective cocktail are shown in the lower panels.}
\label{Fig::Fig2}
\end{figure}

Apart from the signal, the inclusive electron \pt spectrum contains background 
from various sources: conversion of photons including direct photons, Dalitz 
decays of light mesons, dielectron decays of vector mesons, and semileptonic 
decays of kaons ($\rm{K}_{e3}$). The ratio of signal to background ($S/B$) 
depends strongly on \pt. While at low \pt the background dominates the 
inclusive electron yield ($S/B \approx 0.2$ at $\pt = 0.5$~\gevc) the signal
becomes more prominent with increasing \pt ($S/B > 1$ for $\pt > 2.5$~\gevc). 
The background was estimated using a cocktail calculation as described in 
detail in~\cite{Abelev:2012xe}. The main cocktail input is the measured 
\pt-differential production cross section of neutral pions~\cite{ALICE_Pi0}. 
More than 80\% of the electron background can be attributed to $\pi^0$ Dalitz 
decays and the conversion of photons from $\pi^0$ decays. Other light mesons 
($\eta$, $\eta^{\prime}$, $\rho$, $\omega$, $\phi$) were included via 
\mt scaling. About 10\% of the electron background 
at high \pt can be attributed to J/$\psi$ decays. The corresponding cocktail 
input was obtained using a phenomenological interpolation of the J/$\psi$ 
production cross sections measured at various values of \s as described 
in~\cite{Bossu:2011qe}. For direct photons an NLO pQCD calculation was 
used as cocktail input \cite{Gordon:1993qc,Gordon:1994ut}. 
Since the effective material budget was different in the TPC-TOF/TPC and 
TPC-EMCal analysis due to a different requirement on the hits in the SPD 
(Table~\ref{tab::trackcuts}), the amount of background electrons was different 
in the two analyses. In order to estimate the systematic uncertainty of the 
background cocktail, the uncertainties of the various sources were propagated 
in the cocktail as described in~\cite{Abelev:2012xe}. The total systematic 
uncertainty of the cocktail in the TPC-TOF/TPC analysis is smallest at 
\mbox{\pt $\approx$ 1.5 \gevc} where it is \mbox{$\approx$~7\%} and increases 
with increasing \pt reaching 9\% at \mbox{\pt = 4.5 \gevc}. At lower \pt the 
total systematic uncertainty of the cocktail approaches $\approx$ 10\% at 
\mbox{\pt = 0.5 GeV/$c$}. The main contribution comes from the uncertainty on 
the $\pi^{0}$ measurement. In the TPC-EMCal analysis the total systematic 
uncertainty of the cocktail grows from \mbox{$\approx$~9\%} at 
\mbox{$\pt = 4.5$~\gevc} to \mbox{$\approx$~29\%} at 12~\gevc.
The \pt-differential invariant yield of inclusive electrons is compared to
the electron background cocktail in Fig.~\ref{Fig::Fig2} for the
TPC-TOF/TPC analysis (left panel) and the TPC-EMCal analysis (right panel).

The electron background cocktails were statistically subtracted from the 
inclusive electron \pt distributions obtained in the three analyses.
The \pt-differential cross section of electrons from heavy-flavor hadron decays
was obtained by normalizing the invariant yield to the minimum bias cross 
section, which is $55.4 \pm 1.0 ~\rm{mb}$~\cite{abelev:2012sj}. 
The final \pt-differential cross section presented here is a combination of 
the results from the three analyses as summarized in Table~\ref{tab::analyses}. 
In the \pt ranges in which the analyses overlap the results are in agreement 
within their uncertainties. 

\begin{table}
\label{tab::analyses}
\begin{center}
\begin{tabular}{l|c|c|c}
\hline\hline
Analysis & TPC-TOF & TPC & TPC-EMCal \\
\hline
$L_{\rm int}$ (nb)$^{-1}$ & 0.8 & 1.1 & 12.9 \\
\pt range (\gevc) & 0.5 -- 2 & 2 -- 4.5 & 4.5 -- 12 \\
$y$ range         & -0.8 -- 0.8 & -0.8 -- 0.8 & -0.7 -- 0.7 \\
\hline\hline
\end{tabular}
\end{center}
\caption{Integrated luminosities available for the three analyses based on
TPC, TPC-TOF, and TPC-EMCal electron identification, respectively, and
kinematical regions covered by these analyses.}
\end{table}

\section{Results}
\label{sec::Results}
\begin{figure}
  \centering
  \includegraphics[width=0.48\textwidth]{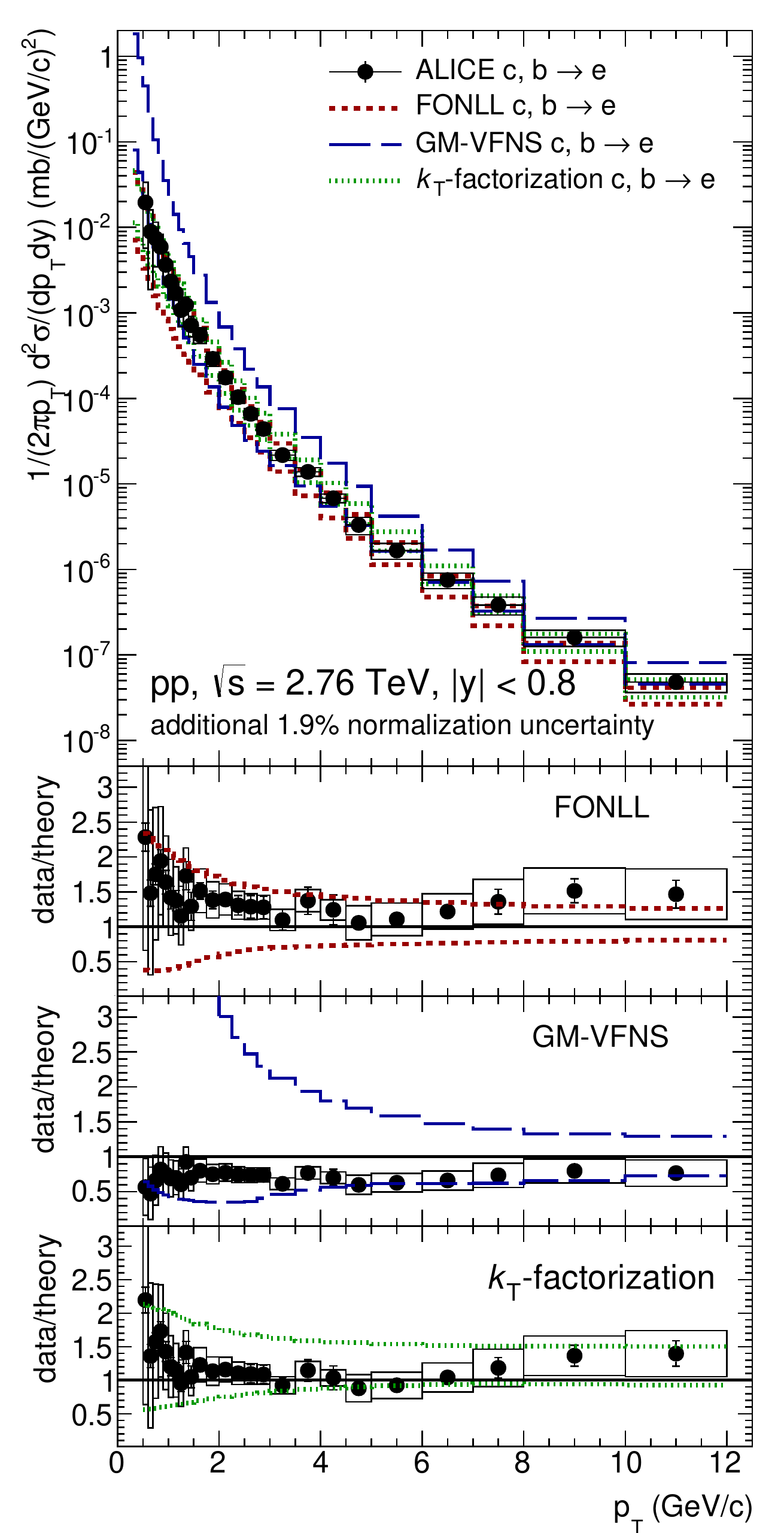}
  \caption{(Color online) \pt-differential cross section of electrons from 
           heavy-flavor hadron decays compared to pQCD calculations from FONLL
           (red)~\cite{fonll,fonll2,Cacciari:2012ny}, GM-VFNS 
           (blue)~\cite{Kniehl:2004fy,Kniehl:2005mk,Kniehl:2012ti,Bolzoni2013253,Bolzoni2013334} 
           and $k_{{\rm T}}$-factorization (green)~\cite{Hagler:2000dda,Baranov:2000gv,Baranov:2004eu,Baranov:2003cd,Jung:2010ey,Jung:2011yt,Kniehl:2008qb,Kniehl:2010iz,Saleev:2012np,Maciula:2013wg}. 
           Uncertainties on the theory calculations originate from the 
           variation of the factorization and the renormalization scales and 
           from the heavy-quark masses. The ratios data/theory are shown in 
           the lower panels, where the dashed lines indicate the additional
           theoretical uncertainties relative to unity.}
  \label{Fig::Fig3}
\end{figure}

The \pt-differential invariant production cross section of electrons from 
heavy-flavor hadron decays at mid-rapidity in \pp collisions at \s = 2.76 TeV 
is shown in comparison to pQCD calculations from 
FONLL~\cite{fonll,fonll2,Cacciari:2012ny}, 
GM-VFNS~\cite{Kniehl:2004fy,Kniehl:2005mk,Kniehl:2012ti,Bolzoni2013253,Bolzoni2013334}, 
and $k_{{\rm T}}$-factorization~\cite{Hagler:2000dda,Baranov:2000gv,Baranov:2004eu,Baranov:2003cd,Jung:2010ey,Jung:2011yt,Kniehl:2008qb,Kniehl:2010iz,Saleev:2012np,Maciula:2013wg} 
in Fig.~\ref{Fig::Fig3}. Statistical and systematic uncertainties of the 
data are shown separately as error bars and boxes, respectively. Dashed lines 
indicate the uncertainties of the pQCD calculations originating from the 
variation of the factorization and normalization scale as well as of the 
heavy-quark masses~\cite{Cacciari:2012ny,Bolzoni2013253,Bolzoni2013334,Maciula:2013wg}. 
As seen in the lower panels of Fig.~\ref{Fig::Fig3}, all pQCD calculations are 
consistent with the measured cross section over the full \pt range within 
combined experimental and theoretical uncertainties. According to the FONLL 
calculation, this range of the electron transverse momentum includes 
approximately 50\% of the charm and 90\% of the total beauty cross section at 
mid-rapidity. The latter contribution starts to dominate from approximately 
4-5~\gevc towards higher transverse momenta. 

\section{Summary}
\label{sec::Summary}
The inclusive differential production cross section of electrons from
charm and beauty hadron decays was measured with ALICE in the transverse 
momentum range \mbox{$0.5 <$\pt$<12$~\gevc} at mid-rapidity in \pp collisions 
at \mbox{\s$= 2.76$~TeV}, which is the same center-of-mass energy as the one
available so far in Pb--Pb collisions at the LHC. pQCD calculations are in good 
agreement with the data. The measurement presented in this article improves 
the reference cross section of electrons from heavy-flavor hadron decays used 
for the measurement of the corresponding nuclear modification factor in 
Pb--Pb collisions, where the current reference is obtained by scaling the 
cross section measured in \pp collisions at \mbox{\s$ = 7$~TeV} to 2.76~TeV 
using FONLL pQCD calculations~\cite{Averbeck:2011ga}.

\newenvironment{acknowledgement}{\relax}{\relax}
\begin{acknowledgement}
\section*{Acknowledgements}
The ALICE Collaboration would like to thank M.~Cacciari, B.A.~Kniehl, G.~Kramer,
R.~Maciu{\l}a, and A.~Szczurek for providing the pQCD predictions for the cross
sections of electrons from heavy-flavor hadron decays. Furthermore, the
ALICE Collaboration would like to thank W.~Vogelsang for providing NLO pQCD
predictions for direct photon production cross sections which were used as
one of the inputs for the electron background cocktail. 
\\
The ALICE Collaboration would like to thank all its engineers and technicians for their invaluable contributions to the construction of the experiment and the CERN accelerator teams for the outstanding performance of the LHC complex.
The ALICE Collaboration gratefully acknowledges the resources and support provided by all Grid centres and the Worldwide LHC Computing Grid (WLCG) collaboration.
The ALICE Collaboration acknowledges the following funding agencies for their support in building and
running the ALICE detector:
State Committee of Science,  World Federation of Scientists (WFS)
and Swiss Fonds Kidagan, Armenia,
Conselho Nacional de Desenvolvimento Cient\'{\i}fico e Tecnol\'{o}gico (CNPq), Financiadora de Estudos e Projetos (FINEP),
Funda\c{c}\~{a}o de Amparo \`{a} Pesquisa do Estado de S\~{a}o Paulo (FAPESP);
National Natural Science Foundation of China (NSFC), the Chinese Ministry of Education (CMOE)
and the Ministry of Science and Technology of China (MSTC);
Ministry of Education and Youth of the Czech Republic;
Danish Natural Science Research Council, the Carlsberg Foundation and the Danish National Research Foundation;
The European Research Council under the European Community's Seventh Framework Programme;
Helsinki Institute of Physics and the Academy of Finland;
French CNRS-IN2P3, the `Region Pays de Loire', `Region Alsace', `Region Auvergne' and CEA, France;
German BMBF and the Helmholtz Association;
General Secretariat for Research and Technology, Ministry of
Development, Greece;
Hungarian OTKA and National Office for Research and Technology (NKTH);
Department of Atomic Energy and Department of Science and Technology of the Government of India;
Istituto Nazionale di Fisica Nucleare (INFN) and Centro Fermi -
Museo Storico della Fisica e Centro Studi e Ricerche "Enrico
Fermi", Italy;
MEXT Grant-in-Aid for Specially Promoted Research, Ja\-pan;
Joint Institute for Nuclear Research, Dubna;
National Research Foundation of Korea (NRF);
CONACYT, DGAPA, M\'{e}xico, ALFA-EC and the EPLANET Program
(European Particle Physics Latin American Network)
Stichting voor Fundamenteel Onderzoek der Materie (FOM) and the Nederlandse Organisatie voor Wetenschappelijk Onderzoek (NWO), Netherlands;
Research Council of Norway (NFR);
Polish Ministry of Science and Higher Education;
National Science Centre, Poland;
 Ministry of National Education/Institute for Atomic Physics and CNCS-UEFISCDI - Romania;
Ministry of Education and Science of Russian Federation, Russian
Academy of Sciences, Russian Federal Agency of Atomic Energy,
Russian Federal Agency for Science and Innovations and The Russian
Foundation for Basic Research;
Ministry of Education of Slovakia;
Department of Science and Technology, South Africa;
CIEMAT, EELA, Ministerio de Econom\'{i}a y Competitividad (MINECO) of Spain, Xunta de Galicia (Conseller\'{\i}a de Educaci\'{o}n),
CEA\-DEN, Cubaenerg\'{\i}a, Cuba, and IAEA (International Atomic Energy Agency);
Swedish Research Council (VR) and Knut $\&$ Alice Wallenberg
Foundation (KAW);
Ukraine Ministry of Education and Science;
United Kingdom Science and Technology Facilities Council (STFC);
The United States Department of Energy, the United States National
Science Foundation, the State of Texas, and the State of Ohio.

\end{acknowledgement}

\bibliographystyle{utphys}
\bibliography{hfe_pp}

\newpage

\appendix
\section{The ALICE Collaboration}
\label{app:collab}



\begingroup
\small
\begin{flushleft}
B.~Abelev\Irefn{org69}\And
J.~Adam\Irefn{org37}\And
D.~Adamov\'{a}\Irefn{org77}\And
M.M.~Aggarwal\Irefn{org81}\And
M.~Agnello\Irefn{org105}\textsuperscript{,}\Irefn{org88}\And
A.~Agostinelli\Irefn{org26}\And
N.~Agrawal\Irefn{org44}\And
Z.~Ahammed\Irefn{org124}\And
N.~Ahmad\Irefn{org18}\And
I.~Ahmed\Irefn{org15}
% ...
\renewcommand\labelenumi{\textsuperscript{\theenumi}~}

\section*{Affiliation notes}
\renewcommand\theenumi{\roman{enumi}}
\begin{Authlist}
\item \Adef{0}Deceased
\item \Adef{idp30130064}{Also at: St. Petersburg State Polytechnical University}
\item \Adef{idp32002352}{Also at: Department of Applied Physics, Aligarh Muslim University, Aligarh, India}
\item \Adef{idp32685104}{Also at: M.V. Lomonosov Moscow State University, D.V. Skobeltsyn Institute of Nuclear Physics, Moscow, Russia}
\item \Adef{idp32928704}{Also at: University of Belgrade, Faculty of Physics and "Vin\v{c}a" Institute of Nuclear Sciences, Belgrade, Serbia}
\item \Adef{idp33234304}{Permanent Address: Permanent Address: Konkuk University, Seoul, Korea}
\item \Adef{idp33774640}{Also at: Institute of Theoretical Physics, University of Wroclaw, Wroclaw, Poland}
\item \Adef{idp34679600}{Also at: University of Kansas, Lawrence, KS, United States}
\end{Authlist}

\section*{Collaboration Institutes}
\renewcommand\theenumi{\arabic{enumi}~}
\begin{Authlist}

\item \Idef{org1}A.I. Alikhanyan National Science Laboratory (Yerevan Physics Institute) Foundation, Yerevan, Armenia
\item \Idef{org2}Benem\'{e}rita Universidad Aut\'{o}noma de Puebla, Puebla, Mexico
\item \Idef{org3}Bogolyubov Institute for Theoretical Physics, Kiev, Ukraine
\item \Idef{org4}Bose Institute, Department of Physics and Centre for Astroparticle Physics and Space Science (CAPSS), Kolkata, India
\item \Idef{org5}Budker Institute for Nuclear Physics, Novosibirsk, Russia
\item \Idef{org6}California Polytechnic State University, San Luis Obispo, CA, United States
\item \Idef{org7}Central China Normal University, Wuhan, China
\item \Idef{org8}Centre de Calcul de l'IN2P3, Villeurbanne, France
\item \Idef{org9}Centro de Aplicaciones Tecnol\'{o}gicas y Desarrollo Nuclear (CEADEN), Havana, Cuba
\item \Idef{org10}Centro de Investigaciones Energ\'{e}ticas Medioambientales y Tecnol\'{o}gicas (CIEMAT), Madrid, Spain
\item \Idef{org11}Centro de Investigaci\'{o}n y de Estudios Avanzados (CINVESTAV), Mexico City and M\'{e}rida, Mexico
\item \Idef{org12}Centro Fermi - Museo Storico della Fisica e Centro Studi e Ricerche ``Enrico Fermi'', Rome, Italy
\item \Idef{org13}Chicago State University, Chicago, USA
\item \Idef{org14}Commissariat \`{a} l'Energie Atomique, IRFU, Saclay, France
\item \Idef{org15}COMSATS Institute of Information Technology (CIIT), Islamabad, Pakistan
\item \Idef{org16}Departamento de F\'{\i}sica de Part\'{\i}culas and IGFAE, Universidad de Santiago de Compostela, Santiago de Compostela, Spain
\item \Idef{org17}Department of Physics and Technology, University of Bergen, Bergen, Norway
\item \Idef{org18}Department of Physics, Aligarh Muslim University, Aligarh, India
\item \Idef{org19}Department of Physics, Ohio State University, Columbus, OH, United States
\item \Idef{org20}Department of Physics, Sejong University, Seoul, South Korea
\item \Idef{org21}Department of Physics, University of Oslo, Oslo, Norway
\item \Idef{org22}Dipartimento di Fisica dell'Universit\`{a} 'La Sapienza' and Sezione INFN Rome, Italy
\item \Idef{org23}Dipartimento di Fisica dell'Universit\`{a} and Sezione INFN, Cagliari, Italy
\item \Idef{org24}Dipartimento di Fisica dell'Universit\`{a} and Sezione INFN, Trieste, Italy
\item \Idef{org25}Dipartimento di Fisica dell'Universit\`{a} and Sezione INFN, Turin, Italy
\item \Idef{org26}Dipartimento di Fisica e Astronomia dell'Universit\`{a} and Sezione INFN, Bologna, Italy
\item \Idef{org27}Dipartimento di Fisica e Astronomia dell'Universit\`{a} and Sezione INFN, Catania, Italy
\item \Idef{org28}Dipartimento di Fisica e Astronomia dell'Universit\`{a} and Sezione INFN, Padova, Italy
\item \Idef{org29}Dipartimento di Fisica `E.R.~Caianiello' dell'Universit\`{a} and Gruppo Collegato INFN, Salerno, Italy
\item \Idef{org30}Dipartimento di Scienze e Innovazione Tecnologica dell'Universit\`{a} del  Piemonte Orientale and Gruppo Collegato INFN, Alessandria, Italy
\item \Idef{org31}Dipartimento Interateneo di Fisica `M.~Merlin' and Sezione INFN, Bari, Italy
\item \Idef{org32}Division of Experimental High Energy Physics, University of Lund, Lund, Sweden
\item \Idef{org33}Eberhard Karls Universit\"{a}t T\"{u}bingen, T\"{u}bingen, Germany
\item \Idef{org34}European Organization for Nuclear Research (CERN), Geneva, Switzerland
\item \Idef{org35}Faculty of Engineering, Bergen University College, Bergen, Norway
\item \Idef{org36}Faculty of Mathematics, Physics and Informatics, Comenius University, Bratislava, Slovakia
\item \Idef{org37}Faculty of Nuclear Sciences and Physical Engineering, Czech Technical University in Prague, Prague, Czech Republic
\item \Idef{org38}Faculty of Science, P.J.~\v{S}af\'{a}rik University, Ko\v{s}ice, Slovakia
\item \Idef{org39}Frankfurt Institute for Advanced Studies, Johann Wolfgang Goethe-Universit\"{a}t Frankfurt, Frankfurt, Germany
\item \Idef{org40}Gangneung-Wonju National University, Gangneung, South Korea
\item \Idef{org41}Gauhati University, Department of Physics, Guwahati, India
\item \Idef{org42}Helsinki Institute of Physics (HIP), Helsinki, Finland
\item \Idef{org43}Hiroshima University, Hiroshima, Japan
\item \Idef{org44}Indian Institute of Technology Bombay (IIT), Mumbai, India
\item \Idef{org45}Indian Institute of Technology Indore, Indore (IITI), India
\item \Idef{org46}Institut de Physique Nucl\'eaire d'Orsay (IPNO), Universit\'e Paris-Sud, CNRS-IN2P3, Orsay, France
\item \Idef{org47}Institut f\"{u}r Informatik, Johann Wolfgang Goethe-Universit\"{a}t Frankfurt, Frankfurt, Germany
\item \Idef{org48}Institut f\"{u}r Kernphysik, Johann Wolfgang Goethe-Universit\"{a}t Frankfurt, Frankfurt, Germany
\item \Idef{org49}Institut f\"{u}r Kernphysik, Westf\"{a}lische Wilhelms-Universit\"{a}t M\"{u}nster, M\"{u}nster, Germany
\item \Idef{org50}Institut Pluridisciplinaire Hubert Curien (IPHC), Universit\'{e} de Strasbourg, CNRS-IN2P3, Strasbourg, France
\item \Idef{org51}Institute for Nuclear Research, Academy of Sciences, Moscow, Russia
\item \Idef{org52}Institute for Subatomic Physics of Utrecht University, Utrecht, Netherlands
\item \Idef{org53}Institute for Theoretical and Experimental Physics, Moscow, Russia
\item \Idef{org54}Institute of Experimental Physics, Slovak Academy of Sciences, Ko\v{s}ice, Slovakia
\item \Idef{org55}Institute of Physics, Academy of Sciences of the Czech Republic, Prague, Czech Republic
\item \Idef{org56}Institute of Physics, Bhubaneswar, India
\item \Idef{org57}Institute of Space Science (ISS), Bucharest, Romania
\item \Idef{org58}Instituto de Ciencias Nucleares, Universidad Nacional Aut\'{o}noma de M\'{e}xico, Mexico City, Mexico
\item \Idef{org59}Instituto de F\'{\i}sica, Universidad Nacional Aut\'{o}noma de M\'{e}xico, Mexico City, Mexico
\item \Idef{org60}iThemba LABS, National Research Foundation, Somerset West, South Africa
\item \Idef{org61}Joint Institute for Nuclear Research (JINR), Dubna, Russia
\item \Idef{org62}Korea Institute of Science and Technology Information, Daejeon, South Korea
\item \Idef{org63}KTO Karatay University, Konya, Turkey
\item \Idef{org64}Laboratoire de Physique Corpusculaire (LPC), Clermont Universit\'{e}, Universit\'{e} Blaise Pascal, CNRS--IN2P3, Clermont-Ferrand, France
\item \Idef{org65}Laboratoire de Physique Subatomique et de Cosmologie, Universit\'{e} Grenoble-Alpes, CNRS-IN2P3, Grenoble, France
\item \Idef{org66}Laboratori Nazionali di Frascati, INFN, Frascati, Italy
\item \Idef{org67}Laboratori Nazionali di Legnaro, INFN, Legnaro, Italy
\item \Idef{org68}Lawrence Berkeley National Laboratory, Berkeley, CA, United States
\item \Idef{org69}Lawrence Livermore National Laboratory, Livermore, CA, United States
\item \Idef{org70}Moscow Engineering Physics Institute, Moscow, Russia
\item \Idef{org71}National Centre for Nuclear Studies, Warsaw, Poland
\item \Idef{org72}National Institute for Physics and Nuclear Engineering, Bucharest, Romania
\item \Idef{org73}National Institute of Science Education and Research, Bhubaneswar, India
\item \Idef{org74}Niels Bohr Institute, University of Copenhagen, Copenhagen, Denmark
\item \Idef{org75}Nikhef, National Institute for Subatomic Physics, Amsterdam, Netherlands
\item \Idef{org76}Nuclear Physics Group, STFC Daresbury Laboratory, Daresbury, United Kingdom
\item \Idef{org77}Nuclear Physics Institute, Academy of Sciences of the Czech Republic, \v{R}e\v{z} u Prahy, Czech Republic
\item \Idef{org78}Oak Ridge National Laboratory, Oak Ridge, TN, United States
\item \Idef{org79}Petersburg Nuclear Physics Institute, Gatchina, Russia
\item \Idef{org80}Physics Department, Creighton University, Omaha, NE, United States
\item \Idef{org81}Physics Department, Panjab University, Chandigarh, India
\item \Idef{org82}Physics Department, University of Athens, Athens, Greece
\item \Idef{org83}Physics Department, University of Cape Town, Cape Town, South Africa
\item \Idef{org84}Physics Department, University of Jammu, Jammu, India
\item \Idef{org85}Physics Department, University of Rajasthan, Jaipur, India
\item \Idef{org86}Physik Department, Technische Universit\"{a}t M\"{u}nchen, Munich, Germany
\item \Idef{org87}Physikalisches Institut, Ruprecht-Karls-Universit\"{a}t Heidelberg, Heidelberg, Germany
\item \Idef{org88}Politecnico di Torino, Turin, Italy
\item \Idef{org89}Purdue University, West Lafayette, IN, United States
\item \Idef{org90}Pusan National University, Pusan, South Korea
\item \Idef{org91}Research Division and ExtreMe Matter Institute EMMI, GSI Helmholtzzentrum f\"ur Schwerionenforschung, Darmstadt, Germany
\item \Idef{org92}Rudjer Bo\v{s}kovi\'{c} Institute, Zagreb, Croatia
\item \Idef{org93}Russian Federal Nuclear Center (VNIIEF), Sarov, Russia
\item \Idef{org94}Russian Research Centre Kurchatov Institute, Moscow, Russia
\item \Idef{org95}Saha Institute of Nuclear Physics, Kolkata, India
\item \Idef{org96}School of Physics and Astronomy, University of Birmingham, Birmingham, United Kingdom
\item \Idef{org97}Secci\'{o}n F\'{\i}sica, Departamento de Ciencias, Pontificia Universidad Cat\'{o}lica del Per\'{u}, Lima, Peru
\item \Idef{org98}Sezione INFN, Bari, Italy
\item \Idef{org99}Sezione INFN, Bologna, Italy
\item \Idef{org100}Sezione INFN, Cagliari, Italy
\item \Idef{org101}Sezione INFN, Catania, Italy
\item \Idef{org102}Sezione INFN, Padova, Italy
\item \Idef{org103}Sezione INFN, Rome, Italy
\item \Idef{org104}Sezione INFN, Trieste, Italy
\item \Idef{org105}Sezione INFN, Turin, Italy
\item \Idef{org106}SSC IHEP of NRC Kurchatov institute, Protvino, Russia
\item \Idef{org107}SUBATECH, Ecole des Mines de Nantes, Universit\'{e} de Nantes, CNRS-IN2P3, Nantes, France
\item \Idef{org108}Suranaree University of Technology, Nakhon Ratchasima, Thailand
\item \Idef{org109}Technical University of Split FESB, Split, Croatia
\item \Idef{org110}The Henryk Niewodniczanski Institute of Nuclear Physics, Polish Academy of Sciences, Cracow, Poland
\item \Idef{org111}The University of Texas at Austin, Physics Department, Austin, TX, USA
\item \Idef{org112}Universidad Aut\'{o}noma de Sinaloa, Culiac\'{a}n, Mexico
\item \Idef{org113}Universidade de S\~{a}o Paulo (USP), S\~{a}o Paulo, Brazil
\item \Idef{org114}Universidade Estadual de Campinas (UNICAMP), Campinas, Brazil
\item \Idef{org115}University of Houston, Houston, TX, United States
\item \Idef{org116}University of Jyv\"{a}skyl\"{a}, Jyv\"{a}skyl\"{a}, Finland
\item \Idef{org117}University of Liverpool, Liverpool, United Kingdom
\item \Idef{org118}University of Tennessee, Knoxville, TN, United States
\item \Idef{org119}University of Tokyo, Tokyo, Japan
\item \Idef{org120}University of Tsukuba, Tsukuba, Japan
\item \Idef{org121}University of Zagreb, Zagreb, Croatia
\item \Idef{org122}Universit\'{e} de Lyon, Universit\'{e} Lyon 1, CNRS/IN2P3, IPN-Lyon, Villeurbanne, France
\item \Idef{org123}V.~Fock Institute for Physics, St. Petersburg State University, St. Petersburg, Russia
\item \Idef{org124}Variable Energy Cyclotron Centre, Kolkata, India
\item \Idef{org125}Vestfold University College, Tonsberg, Norway
\item \Idef{org126}Warsaw University of Technology, Warsaw, Poland
\item \Idef{org127}Wayne State University, Detroit, MI, United States
\item \Idef{org128}Wigner Research Centre for Physics, Hungarian Academy of Sciences, Budapest, Hungary
\item \Idef{org129}Yale University, New Haven, CT, United States
\item \Idef{org130}Yonsei University, Seoul, South Korea
\item \Idef{org131}Zentrum f\"{u}r Technologietransfer und Telekommunikation (ZTT), Fachhochschule Worms, Worms, Germany
\end{Authlist}
\endgroup

\end{document}